# Health Care Crowds: Collective Intelligence in Public Health

JOHN PRPIĆ – Faculty of Business Administration, Technology & Social Sciences, Lulea University of Technology

## 1.0 INTRODUCTION

For what purposes are crowds being implemented in health care? Which crowdsourcing methods are being used? This work begins to answer these questions by reporting the early results of a systematic literature review of 110 pieces of relevant research. The results of this exploratory research in progress reveals that collective intelligence outcomes are being generated in three broad categories of public health care; health promotion, health research, and health maintenance, using all three known forms of crowdsourcing. Stemming from this fundamental analysis, some potential implications of the research are discussed and useful future research is outlined.

## 2.0  DATA COLLECTION & FILTERING

Data collection through the use of secondary archival sources such as search engines, alerts, social media, web pages, the general press, blogs, and the research literature itself, began in October 2013 and continues through this research in progress. Thus far, 153 pieces of literature have been collected, 110 of which remained relevant for this work after data filtering. The 110 articles captured in this analysis are limited to articles that research the use of crowdsourcing in the health area, and all the research works in question readily self-identify in this manner, to greater or lesser extent. Of note, the considerable literature on Citizen Science (Newman et al 2012, Wiggins and Crowston 2011) has been completely excluded in the data reviewed here. Though Citizen Science as a practice also uses crowds to generate collective intelligence, specifically in the guise of scientific research, such work is not directly for healthcare, and thus it is excluded in this analysis. Further, in the context of this work, literature is inclusive of the following types only; journal articles, conference articles, journal & conference editorials, and theses.

To give the reader some indication of the depth of the search for literature, in early 2015 a Google Scholar search was undertaken for "crowdsourcing"+"health", and 35 pages of the search results (ie the first 600 results) were reviewed entry by entry the investigator until search saturation was achieved; indicated by the duplication of previous entries and a concomitant lack of new relevant results. Altogether, the set of literature considered, though extensive, is unlikely to be similarly comprehensive, and so the findings of this work should be viewed as nothing more than a solid platform for further work.

## 3.0 ANALYTICAL LENS

One perspective on collective intelligence views it as resulting from different forms of IT-mediated crowdsourcing (Prpić et al 2014, de Vreede et al. 2013). In this work, we employ this typology as our analytical lens to organize our analysis of health care crowds. More specifically, a *virtual labor marketplace* is an IT-mediated market for spot/piecework labor (Kingsley et al 2014), where individuals and organizations can agree to execute work in exchange for monetary compensation. This variety of IT-mediated collaboration is exemplified by endeavors like Crowdflower and Amazon's M-Turk (Cheng et al 2015, Horton 2010).  In *tournament-based collaboration* organizations post their problems/opportunities to IT-mediated crowds at web properties such as Innocentive, Kaggle, and eYeka. In posting a problem/opportunity at these intermediaries, the organization creates a competition amongst the crowd assembled there, where the best solution will be chosen as determined by the organization (Boudreau & Lakhani 2013, Afuah & Tucci 2012). In an *open collaboration* model, organizations post their problem to the public at large through IT. Contributions from crowds in these endeavors are voluntary and thus, do not generally entail monetary exchange. Employing a wiki or using social media for input, are examples of this type of collaboration (Sutton et al 2014, Crump 2011). In sum, each piece of research in the set of literature analysed below has been coded as falling into one of the three categories of crowdsourcing modality discussed here.

## 4.0 ANALYSIS

Building on the recent work of Parvanta et al (2013), Ranard et al (2013), Steele (2013), Brabham et al (2014) and Ghosh & Sen (2015), this work explores the refined set of literature gathered, and from said exploration infers that the literature pertaining to the use of crowds in health care can be found to fall into the following three categories:

- Health Promotion
- Health Research
- Health Maintenance

In this context, health promotion represents all the literature pertaining to the use of or discussion of IT-mediated crowds for the promotion of public health care, and includes health care activities such as; disease detection (Susumpow et al 2014, Bodnar & Salathe' 2013) and surveillance (Chunara et al 2013, Boulos et al 2011), behavioural interventions (Morris et al 2011, Noronha et al 2011), health literacy (Johansson et al 2013, Rubenstein 2013), and health education (Moskowitz et al 2015, Tuominen et al 2014). Moreover, the health research category encompasses literature pertaining to the use or discussion of IT-mediated crowds for public health research, and includes activities such as; pharmaceutical research (Adams 2014, Ekins & Williams 2010), clinical trials (Darrow 2014, Kuffner et al 2014) health experiment methodology (Barsnes & Martens 2013, Schmidt 2010), building and improving health care research knowledge (Mortenson et al 2013, McCoy et al 2012). Similarly, the health maintenance category encompasses literature pertaining to the use or discussion of IT-mediated crowds for health care treatment, and includes activities such as; patient-related (Vennik et al 2014, Mentha & Toso 2014), physician-related (Cheng et al 2015, Sims et al 2014), diagnostics (Tucker et al 2014, Freifeld et 2010), medical practice (Pradhan & Gay 2014, Merchant et al 2013), and treatment support (Griffiths et al 2014, Amir et al 2013). Taken together, these three categories are both exhaustive and exclusive, in the sense that all 110 pieces of literature considered here fall into only one category. In Table #1 below, the relative distribution of each piece of literature in each health category is illustrated, organized by the specific form of Crowdsourcing that the literature discusses.

**Table #1 – Distribution of Literature by Type of Crowdsourcing & Health Care Category**

| Health Crowd Category | Health Promotion (36 pieces of literature) | Health Research (39 pieces of literature) | Health Maintenance (35 pieces of literature) |
|---|---|---|---|
| **Form of Crowdsourcing** | | | |
| Virtual Labor-Markets | 1 | 8 | 1 |
| Tournament-Based Collaboration | 2 | 4 | 2 |
| Open Collaboration | 33 | 27 | 32 |

## 5.0 DISCUSSUION

When considering Table #1, it appears that the three health care categories proposed here cluster into relatively equal values of 35-40 pieces of literature each. Further, though approximately 5/6 of the dataset discusses open collaboration crowdsourcing in particular, it's notable that all forms of crowdsourcing are represented in each category of health care. This would seem to indicate that in some respect, all forms of crowdsourcing can bring value to different areas of public health care concern. Similarly, the health research category shows both the greatest amount of overall literature, and the greatest diversity of crowdsourcing methods employed, which may indicate that the use of IT-mediated crowds for health research is the most advanced crowdsourcing sub-field in the public health domain.

As it now stands in this analysis, the use of open collaboration crowdsourcing techniques dominates the conversation and application of IT-mediated crowds for public health care, and the reason(s) for this are unclear. It may be that open collaboration applications such as social media, wikis, and collaborative mapping are the most mature applications overall, thus, being more widespread and well-known, and hence, explaining the predominance in the data sample. On the other hand, it may be that open collaboration crowdsourcing techniques are the most suited or most useful for public health use, thus explaining their predominance. More research will be needed to distinguish whether these early returns are due to the former or the latter.  Similarly, in this analysis the rather widespread use of open collaboration crowdsourcing in all three areas of public health care needs further comparison amongst the health care categories. Said another way, what can researchers and practitioners in each health care category using open collaboration crowdsourcing (or the other methods) learn from one another?

Further, this work focusses on the research literature at the expense of health crowd phenomena 'in the wild', and so, more data is needed on the specific applications being used or proposed for the public health use of IT-mediated crowds. Along these lines, recent work by Prpić et al (2014b) vividly illustrates the relative trade-offs inherent to the different forms of crowdsourcing, and thus further work delineating these trade-offs in public health care may be very useful.

## 6.0 CONCLUSION

This work reports the early results of exploratory research in progress, which begins to unpack the purposes for which IT-mediated crowds are being implemented in public health care, and the crowdsourcing methods being used therein, stemming from a systematic literature review of 110 pieces of relevant research. The results of this exploratory research reveal that collective intelligence outcomes are being generated in three broad categories of public health care; health promotion, health research, and health maintenance, using all three known forms of crowdsourcing. In this respect, this work contributes to the public health care literature by forming an extensive dataset of literature directly relevant to the use of crowdsourcing in this important domain, and in doing so we learn that all methods of crowdsourcing are being used in all three broad areas of public health concern. Similarly, we learn that the use of open collaboration is far and away the most prevalent method of crowdsourcing in the public health domain. Further, this work contributes to the corpus of the crowdsourcing literature by unpacking and organizing the emerging use of crowdsourcing techniques in an important new domain of application, illustrating that these techniques are both viable and valuable for largely non-commercial uses. Altogether, it is hoped that this fundamental work forms a solid platform for future inquiry, benefitting researchers and practitioners alike.